\documentclass[12pt,preprint]{aastex}
\usepackage{times}
\usepackage{graphics,epsf}
\usepackage{amsmath}                
\usepackage{amsfonts}               
\usepackage{amssymb}                
\usepackage{epsfig}                 
\usepackage{rotating}
\usepackage{color}
\usepackage{tabularx}
\usepackage{upgreek}

\def \cm{~\rm{cm}}
\def \s{~\rm{s}}
\def \km{~\rm{km}}

\def \K{~\rm{K}}
\def \g{~\rm{g}}

\def \erg{~\rm{erg}}

\def \yr{~\rm{yr}}

\def \dif{\mathrm{d}}
\def \apj{ApJ}
\def \aap{A\&A}

\def \mnras{MNRAS}
\def \apjl{ApJ Lett.}
\def \apjs{ApJ Suppl. Ser.}

\def \pasp{PASP}
\def \aapr{A\&A Rev}
\def \araa{ARA\&A}
\def \na{New A}

\begin{document}

\title{Ejecting the envelope of red supergiant stars with jets launched by an inspiraling neutron star}

\author{Oded Papish\altaffilmark{1}, Noam Soker\altaffilmark{1} and Inbal Bukay\altaffilmark{1}}

\altaffiltext{1}{Department of Physics, Technion -- Israel Institute of Technology, Haifa
32000, Israel; papish@techunix.technion.ac.il; soker@physics.technion.ac.il;inbalbukay@gmail.com}

\begin{abstract}

We study the properties of jets launched by a neutron star
spiralling inside the envelope and core of a red supergiant.  We
propose that Thorne--\.{Z}ytkow objects (TZO) are unlikely to be
formed via common envelope (CE) evolution if accretion on to the
neutron star can exceed the Eddington rate with much of the
accretion energy directed into jets that subsequently dissipate
within the giant envelope. We use the jet-feedback mechanism,
where energy deposited by the jets drives the ejection of the
entire envelope and part of the core, and find a very strong
interaction of the jets with the core material at late phases of
the CE evolution. {{{{{{ Following our results we speculate on two
rare processes that might take place in the evolution of massive
stars. (1)  Recent studies have claimed that the peculiar
abundances of the HV2112 RSG star can be explained if this star is
a TZO. We instead speculate that the rich-calcium envelope comes
from a supernova explosion of a stellar companion that was only
slightly more massive than HV2112, such that during its explosion
HV2112 was already a giant that intercepted a relatively large
fraction of the SN ejecta.} }}}}} (2)  We raise the possibility
that strong r-process nucleosynthesis, where elements with high
atomic weight of $A \ga 130$ are formed,
  occurs inside the jets that are launched by the NS inside the core of the RSG star.

\end{abstract}

\section{INTRODUCTION}
\label{sec:intro}

{{{{{ Massive stars, in particular those in interacting binary
systems, hold many secretes in their evolution. Some of these
puzzles are related to the synthesis of different isotopes. Our
study is related to the peculiar abundances of some isotopes in
the red supergiant (RSG) star HV2112 in the Small Magellanic
Cloud, and to the site of the strong r-process. The paper is
centered around jets that are assumed to be launched by the
neutron star (NS) as it accretes material from the common envelope
(CE) with a RSG star.

Contrary to recent claims that HV2112 is a Thorne-\.{Z}ytkow
object (TZO; \citealt{Levesqueetal2014, Toutetal2014}), we show
that a TZO is unlikely to be formed through the evolution of a NS
inside the envelope of a RSG star. The reason is that the jets
launched by the NS expel the entire envelope and most of the core.
}}}}}

{{{{{ Another major problem in astrophysics is the }}}}} exact
sites where the r-process nucleosynthesis takes place. For the
site(s) of the ``strong r-process'', where elements of atomic
weight $A \ga 130$ are formed, there have been two main contenders
in recent years (e.g., \citealt{Thielemannetal2011}): the merge of
two neutron stars (e.g., \citealt{Qian2012, Rosswogetal2013}, and
references therein), and jets from a rapidly rotating newly
born single NS \citep{Winteleretal2012}.

Previous studies of r-process nucleosynthesis in jets include
\citet{Cameron2001} who suggested the possibility of creating
r-process elements inside the jets launched at a velocity of
$\frac{1}{2} c$ from an accretion disk around a rapidly rotating
proto-NS. \citet{Nishimura2006} simulated the r-process
nucleosynthesis during a jet-powered explosion. In their
simulation a rotating star with a magnetic field induced a
jet-like outflow during the collapse which explodes the star.
Neutrinos played no role in their simulation. \cite{PapishSoker2012}
calculated the nucleosynthesis inside the hot bubble formed in the
jittering-jets model for core-collapse supernova explosions (CCSN), and
found that substantial amount of r-process material can be formed.
\cite{PapishSoker2012} assumed that in the jittering-jets
explosion model the jets are launched close to the NS where the
gas is neutron-rich (e.g., \citealt{kohri2005}).

Some r-process elements are believed to be found in all
low-metallicity stars, and thus r-process nucleosynthesis must
take place continuously from very early times in the Galactic evolution
\citep{Sneden2008}. Since the jittering-jets model was constructed
to account for the explosion of {\it all} CCSNe, the r-process induced by the jittering jets can account
for the continuous formation of r-process elements. However, the
abundance of heavy r-process elements has much larger variations
among different stars, implying that part of the heavy r-process,
termed strong r-process, is formed in rare events
(\citealt{Qian2000,Argast2004}).

Here we propose a new {speculative} site for strong r-process
nucleosynthesis in which it is formed by jets launched by a NS
spiraling-in inside the core of a giant star. These jets will also
explode the star. This is a rare evolutionary route and hence
complies with the finding of large variations in the abundances of
these elements. In this first study we limit ourselves to present
the scenario and show its viability. Most ingredients of our newly
proposed scenario were studied in the past but were never put
together into a coherent picture to yield a new possible site for
the strong r-process. Previously studied ingredients of our
proposed scenario include the CE of a NS and a giant
(\citealt{ThorneZytkow1975, Armitage2000, Chevalier2012}), the
launching of neutron-rich gas from accretion disks around compact
objects (e.g., \citealt{Surman2004,kohri2005}), the launching of
jets by NS accreting at a high rate \citep{Fryer1996}, and the
formation of r-process elements in jets from NS \citep{Fryer2006}.
The idea that jets can explode stars {{{ under specific conditions
}}} was also studied in the past, e.g., in a CE evolution
\citep{Chevalier2012}. {{{ Another specific class of models are
based on magnetic amplification by a rapidly rotating stellar core
}}}  (e.g. \citealt{LeBlanc1970, Meier1976, Bisnovatyi1976,
Khokhlov1999, MacFadyen2001, Woosley2005,Couch2009}). {{{ This
magnetorotational mechanism creates bipolar outflows (jets) around
the newly born NS that are able to explode the star. However, the
required core's rotation rate is much larger than what stellar
evolution models give, hence making most of these models
applicable for only special cases. }}} We, on the other hand,
claim that {\it all} CCSNe are exploded by jets, the
jittering-jets model, that also synthesize r-process elements (but
not the strong r-process), and hence our newly proposed scenario
is part of a unified picture we try to construct for exploding
{\it all } massive stars and the synthesis of r-process elements.
{{{ In the jittering-jets model the explosion of CCSNe is powered
by jittering jets launched by an intermittent accretion disk
around the newly born NS \citep{Papishsoker2014,
Papishsoker2014b}. The intermittent accretion disk is formed by
gas accreted from the convective core regions that have a
stochastic angular momentum \citep{Papishsoker2014,
GilkisSoker2014}. }}}

\citet{Taam1978} already studied the spiraling-in of a NS inside a
red supergiant envelope and considered two outcomes, that of
envelope ejection, and that of core-NS merger. However, they did not
consider jets. The removal of the envelope during a CE evolution
by jets launched by a NS or a WD companion was discussed by
\cite{Armitage2000} and \cite{Soker2004}. \cite{Chevalier2012}
explored the possibility that the mass loss prior to an explosion
of a core-NS merger process is driven by a CE evolution of a NS
(or a BH) in the envelope of a massive star. {{{{{ As already
discussed by \cite{Armitage2000}, }}}}} jets launched by an
accretion disk around the NS companion deposit energy to the
envelope and help in removing the envelope. WDs can do the same
\citep{Soker2004}. However, for a NS the accretion process to form
an accretion disk is much more efficient due to neutrino cooling
\citep{HouckChevalier1991, Chevalier1993, Chevalier2012}. This
allows the NS to accrete at a very high rate, much above the
classical Eddington limit, hence leading to an explosive
deposition of energy to the envelope \citep{Chevalier2012}. The
process by which jets launched by an inspiraling NS expel the
envelope \citep{Armitage2000, Soker2004} and then the core in an
explosive manner \citep{Chevalier2012} implies that no
Thorne-\.{Z}ytkow objects (\citealt{ThorneZytkow1975}; a red giant
with a NS in its core) can be formed via a CE evolutionary route.
We strengthen this conclusion in section \ref{sec:commonenvelope}.

In section \ref{sec:commonenvelope} we study the CE evolution
toward the explosion and estimate the typical accretion rate and
time scale. A new ingredient in our CE calculation is the
assumption that mainly jets launched by the inspiraling NS drive
the envelope ejection via the jet-feedback mechanism
\citep{Soker2004, Sokeretal2013, Soker2014}. It is possible that
the NS is able to expel the entire envelope by the jets
\citep{Soker2004}, and as a consequence cannot enter inside the core. However, it seems
that a merger at the end of the CE phase can be quite common
\citep{Soker2013}, and we study such cases. {{{{{ In section
\ref{sec:implciaitons} we list two, somewhat speculative,
processes that are motivated by our results.  }}}}} Our summary is
in section \ref{sec:summary}.

\section{JETS INSIDE A COMMON ENVELOPE}
\label{sec:commonenvelope}
\subsection{General derivation}
\label{sec:derivation}
The Bondi-Hoyle-Lyttleton (BHL) mass accretion rate {{{
\citep{HoyleLyttleton1939, BondiHoyle1944} }}} inside the envelope
is (e.g., \citealt{Armitage2000}; for a general overview of CE see
\citealt{Ivanovaetal2013})
\begin{equation}
\dot M_{\rm BHL}=\uppi
\left(\frac{2GM_{NS}}{v_{r}^2+c_{s}^2}\right)^2 \rho_{e}
\sqrt{v_{r}^2+c_{s}^2 },
\end{equation}
where $M_{\rm NS}$ is the NS mass, $v_{r}$ its velocity relative
to the giant's envelope, $c_{s}$ is the sound speed inside the
envelope, and $\rho_{e}$ is the envelope density at the NS
location. The NS moves mildly supersonically and we can use the
approximation $\sqrt{v_{r}^2+c_{s}^2 } \simeq v_{K}$, where
$v_{K}$ is the NS's Keplerian velocity.

We use the red giant's profile from \citet{Taam1978} with a mass
of $M_{g}=16M_\odot$, a radius of $R_{g}=535R_\odot$, a core
radius of $R_{\rm core}=2.07\times 10^{10} \cm$, and a core mass
of $M_{\rm core}=7\times 10^{33} \g \simeq 3.5M_\odot$. A power
law fit of the density profile in the giant's  envelope gives
\begin{equation}
\rho _{e} \simeq Ar^\beta = 0.68 \left( \frac{r}{R_\odot} \right)^{-2.7} \g \cm^{-3},
\end{equation}
where $r$ is the radial coordinate of the star.
The BHL accretion rate can then be written as
\begin{eqnarray}
\label{eq:m_bh}
\dot{M}_{\rm BHL} \cong \uppi \left(\frac{2GM_{\rm NS}}{v_{K}^2}\right)^2\rho_{e}v_{K}=4\uppi a^2\left(\frac{M_{NS}}{M(a)}\right)^2\rho_{e}v_{K}  \nonumber
\\
\simeq 3 \times 10^3 \left( \frac{a}{1R_\odot} \right)^{-1.2}
\left( \frac{M(a)}{7 M_\odot} \right)^{-3/2}
M_\odot \yr^{-1},
\end{eqnarray}
where $a$ is the distance between the NS and the core's center, $M_{\rm g}(r)=M_{\rm core}+M_{\rm env}\left( r/R_{\rm g}\right)^{0.3}$
is the giant's mass enclosed inside radius $r$, $M(r)= M_{\rm g}(r) + M_{\rm NS}$, and we take $M_{\rm NS}=1.4 M_\odot$ in the rest of the paper.
We take the mass accretion rate to be a fraction $f_a$ of the BHL accretion rate, and we take a fraction $\eta \simeq 0.1$ of the accreted mass to be launched in the two jets
\begin{equation}
\dot M_{2\rm j}=\eta \dot M{\rm acc}= \eta f_a \dot M_{\rm BHL}.
\label{eq:macc0}
\end{equation}

\begin{figure}[h!]
\begin{center}
\includegraphics[width=0.3\textwidth]{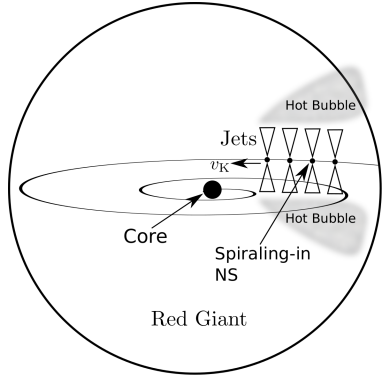}
\caption{A schematic drawing (not to scale) of the spiral-in process while the jets release energy. As the NS spirals-in,
the jets' heads encounter different parts of the envelope, and release their energy at different regions.
A continuous  hot bubble is formed in each side of the equatorial plane \citep{Sokeretal2013}.}
\label{fig:spiral-in}
\end{center}
\end{figure}

{{{{Around the NS an accretion disk is easily formed as the ratio
of orbital separation to the NS radius is very large, such that
the specific angular momentum of the accreted gas is more than an
order of magnitude above what is required to form an accretion
disk (eq. 7 of \citealt{Soker2004}).} }}} Such an accretion disc
can launch two bipolar jets inside the envelope as shown
schematically in Fig. \ref{fig:spiral-in}. As a result of the NS
motion the jets will encounter different parts of the envelope
during their propagation and form a hot bubble on each side of the
orbital plane \citep{Sokeretal2013}.

To examine the fate of the jets we compare the propagation time
scale of the jets inside the envelope $\tau_j$ with the time scale
it takes for a jet's head to cross its width $\tau_c$. $\tau_j$
can be estimate as follows. The jets' head velocity $v_{h}$ is
determined by the pressure balance on the two sides of the jets'
head $\rho_{e} v_{h}^{2}\simeq \rho_{j}(v_{j}-v_{h})^{2} \simeq
\rho_j v_j^2$, as $ v_{h}\ll v_{j}$; $v_j$ is the initial velocity
of the gas in the jet, {{{ about the escape speed from the neutron
star, }} } and $\rho_e$ is the envelope density encountered by the
jet. We take the jet to have a half opening angle of $\theta
\approx 0.2$ {{{ (the exact value of $\theta$ is not important, it
can be any value less than about 0.5), }}} and a mass outflow rate
into each jet of $\frac{1}{2} \eta \dot M_{\rm acc}$. The jets'
head velocity is then
\begin{equation}
v_h(z) \simeq v_j\sqrt{\frac{\rho_{j}}{\rho_{e}(z)}} =
\sqrt{\frac{\frac{1}{2} \eta \dot M_{\rm acc}v_j}{\uppi z^{2}
\theta^{2}A(a^{2}+z^{2})^{\beta/2}}},
\end{equation}
where we take the jets' density to be
\begin{equation}
\rho_{j}=\frac{\frac{1}{2}\eta \dot M_{\rm acc}}{\uppi z^{2} \theta^{2} v_{j}},
\end{equation}
as $\theta \ll 1$. The propagation time of the jets to a distance
$z=\int^{\tau_j}_0 v_h {\rm{d}} t$ can then be solved analytically
\begin{equation}
\tau_j = \frac{a^{2+\beta/2}}{2+\beta/2}\theta\sqrt{\frac{2A\uppi}{\eta\dot M_{\rm acc}v_j}}\left[ \left(1 +\frac{z^2}{a^2}\right)^{1+\beta/4}-1\right].
\end{equation}

We substitute $\dot M_{\rm acc}$ from equation \ref{eq:macc0}, take $\beta= -2.7$, define $v_{j5} \equiv v_j/10^5 \km \s^{-1}$,
and scale quantities for the inner part of the envelope to get
\begin{eqnarray}
& \tau_j \simeq 3 \times 10^3 v_{j5}^{-1/2} \left( \frac{a}{R_\odot}\right)^2 \left( \frac{\theta}{10^\circ}\right)
\left( \frac{\eta}{0.1}\right)^{-1/2}   \nonumber
\\
&\times \left( \frac{M_{\rm NS}}{1.4 M_\odot} \right) \left( \frac{M(a)}{7 M_\odot} \right)^{5/4}
\left[ \left(1 +\frac{z^2}{a^2}\right)^{1+\beta/4}-1\right] \s.
\end{eqnarray}
For the crossing time we take \citep{Soker2004}
\begin{equation}
\tau_c=\frac{2 z\theta}{v_{K}} \simeq 200 \left( \frac{\theta}{10^\circ}\right)  \left( \frac{z}{R_\odot}\right)  \left( \frac{a}{R_\odot}\right)^{1/2}
\left( \frac{M(a)}{7 M_\odot}\right) \s.
\end{equation}
The ratio between these times is given by (note that $\theta$ cancels out)
\begin{eqnarray}
& \frac{\tau_j}{\tau_c} \simeq 15 v_{j5}^{-1/2} \left( \frac{a}{R_\odot}\right)^{1/2} \left( \frac{a}{z}\right)  \left( \frac{\eta}{0.1}\right)^{-1/2}
   \nonumber
\\
& \times \left( \frac{M_{\rm NS}}{1.4 M_\odot} \right)  \left( \frac{M(a)}{7 M_\odot} \right)^{1/4}
\left[ \left(1 +\frac{z^2}{a^2}\right)^{1+\beta/4}-1\right]  .
\end{eqnarray}
The jets are {{{{unlikely} }}}to penetrate through the envelope when $\tau_c
< \tau_j$. {{{ The reason is as follows. The moment a fresh supply
of jet material ceases along a particular direction, the
propagation of the jet's head stops along that direction. The
inequality $\tau_c < \tau_j$ implies that before the jet's head
manages to exit the star along any direction, a fresh supply of
jet material has ceased; the jet cone has crossed that direction
and moved to other directions. }} } {{{{ This process was
simulated in a very simple preliminary setting
\citep{Sokeretal2013}, but it requires detail numerical
simulations to determine the exact outcome, and under what
conditions the jets dissipate in the envelope. From simulations of
active galactic nuclei jets we know that a relative motion of the
jets' source and the inter-stellar medium leads to dissipation and
bubble inflation \citep{Sokeretal2013}. }}}}

For $z \approx a$, $\beta = -2.7$ and $\eta = 0.1$ the ratio is
\begin{equation}
\frac{\tau_j}{\tau_c} \simeq 4 \left( \frac{a}{R_\odot}\right)^{1/2}.
\end{equation}
We find that the condition $\tau_c < \tau_j$ is satisfied.
We conclude that the jets {{{{may}  }}} not be able to penetrate through the envelope and will deposit their energy inside the envelope.

The total energy deposited by the jets inside the envelope is
\begin{equation}
E_{j} =\int \dot E_{j} {\rm d}t = \int \frac{1}{2} \eta \dot
M_{\rm acc} v_j^2 ~\dif t,
 \label{ejej}
\end{equation}
{{{{ where integration is over the lifetime of the jets. }}}}
 $\dot M_{\rm acc}$ can be eliminated with an expression that equates the power of the interaction of the NS with the envelope, i.e.,
 the work of the drag force,  with the releasing rate of orbital energy \citep{Iben1993}
\begin{equation}
 -\frac{GM_{\rm g}(a)M_{\rm NS}}{2a^2} \dot a= \xi \dot M_{\rm BHL} v_{K}^2,
\label{L}
\end{equation}
where $\xi \simeq 1$ is a parameter characterizing the drag force
(accretion + gravitational interaction with the rest of the
envelope). Equations (\ref{ejej}) and (\ref{L}) give
\begin{equation}
\int \dif E_j = -\int _{R_g} ^{a} \frac{1}{4a^\prime} \frac {\eta
f_a}{\xi} \frac{M_{\rm g}(a^\prime)M_{\rm NS}}{M(a^\prime)} v_j^2
\dif a^\prime,
\end{equation}
where $R_g$ is the giant radius, and $a$ is the final location of
the NS. Approximating the reduced mass by $M_g(a) M_{\rm NS}/M(a)
\simeq M_{\rm NS} $ allows us to perform the integration
analytically to obtain
\begin{equation}
E_j \simeq \frac{1}{4} \frac {\eta f_a}{\xi}  v_j^2 M_{\rm NS} \left[ \ln\left(\frac{R_{\rm g}}{a}\right) \right].
\label{eq:ej}
\end{equation}
As the jets {{{ are unlikely to }}} penetrate the envelope, we assume that there is a self regulation process, i.e., a  negative feedback process \citep{Sokeretal2013}, such that the jets from the NS prevent the
accretion rate on to the NS from being too high.

\subsection{Jets inside a common envelope}
\label{sec:cenvelope}

{{ { Few words on the nature of the jet launching setting are in
place here. First, we note that some recent numerical simulations
of CE evolution have showed that there is an initial rapid phase
of evolution in which the orbit is shrunk significantly while the
outer parts of the CE are inflated (e.g., \citealt{Ricker2012,
DeMarcoetal2012, Passyetal2012}). Even in that phase the accretion
rate by the companion can be quite high, despite being below the
BHL rate. In the simulation of \cite{Ricker2012} the rapid
plunge-in phase lasts for about 20 days, and the accreted mass
during that period is $M_{\rm acc} \simeq 0.0003 M_\odot$, giving
an accretion rate of $\dot M_{\rm acc} \simeq 0.005 M_\odot
\yr^{-1}$. This is more than what is required for neutrino cooling
to allow high accretion rate \citep{Chevalier1993}. So the NS can
accrete at a very high rate along the entire CE evolution of the
cases studied here. }

{ Another point to emphasize is that the accretion energy can in
principle be channelled to neutrinos, jets, and electromagnetic
(EM) radiation. We take the view that at very high accretion
rates, both on to NS and main sequence stars, a large fraction of
the energy is carried by jets rather than by EM radiation
(\citealt{Soker2015} and references therein). The neutrino cooling
in the present setting allows a very high accretion rate, but we
assume that the jets' power is much above the Eddington luminosity
limit. The jets are collimated, such that accretion can proceed
from directions perpendicular to the jets' axis. In this process
most of the energy in the disk that is not carried by neutrinos is
transferred to magnetic fields that by violent reconnection eject
mass. Namely, energy is channelled to magnetic fields and outflows
much more than to thermal energy and EM radiation (Shiber, S.,
Schreier, R., \& Soker, N., 2015, in preparation). The jets are
not only a byproduct of the high accretion rate allowed by
neutrino cooling, but by themselves allow an even higher accretion
rate. }

{ Thirdly, we point out that part of the evolution might take
place in a grazing envelope evolution (GEE) rather than a CE
evolution. In the GEE a stellar companion (NS or a main sequence
star) grazes the envelope of a giant star while both the orbital
separation and the giant radius shrink simultaneously
\citep{Soker2015}. The orbital decay itself is caused by the
gravitational interaction of the secondary star with the envelope
inward to its orbit, i.e., dynamical friction (gravitational
tide). The binary system might be viewed as evolving in a constant
state of `just entering a CE phase'. The GEE is made possible only
if the companion manages to accreted mass at a high rate and
launch jets that remove the outskirts of the giant envelope, hence
preventing the formation of a CE. This might occur when the NS is
in the low density parts of the envelope. Eventually it enters the
dense part of the core (see below), and a CE phase commences.
Hence, the occurrence of a GEE in the outer parts of the envelope
does not change our conclusions. }
  }}

\subsection{Inside the envelope}
\label{sec:envelope}
Let the removal of the envelope by the jets have an efficiency of $\chi$, such that
\begin{equation}
\chi E_j \la E_{\rm e/bind}.
\label{eq:chi}
\end{equation}
The removal energy supplied by the jets is smaller than the
binding energy of the envelope since in addition to the energy
deposited by the jets there is the orbital energy of the
spiraling-in binary system. In the traditional CE calculation only
the orbital energy is considered. The binding energy of the
envelope of the model described in section \ref{sec:derivation} is
\begin{equation}
E_{\rm e/bind} \simeq \frac{ G M_{\rm env}^2 }{R_{\rm core}} \left[
\frac{3}{7} \frac{M_{\rm core}}{M_{\rm env}}
\left( \frac{R_{\rm core}}{R_g} \right)^{0.3}
+
\frac{3}{4}
\left( \frac{R_{\rm core}}{R_g} \right)^{0.6} \right].
\label{eq:e_bind}
\end{equation}
For $R_{\rm core} \simeq 0.3 R_\odot$, $M_{\rm core} =  3.5 M_\odot$ and $M_{\rm env}= 12.5 M_\odot$ we obtain
$ E_{\rm e/bind} \simeq 4 \times 10^{49} \erg$.
Substituting equations (\ref{eq:ej}) and (\ref{eq:e_bind}) in
equation (\ref{eq:chi}) gives
\begin{equation}
f_a \la 0.01 ~\xi \left(\frac{\eta}{0.1}\right)^{-1} \left(\frac{\chi}{0.01}\right)^{-1}.
\end{equation}
Substituting typical values in equation (\ref{eq:m_bh}) with $\chi
\simeq 0.01$ we find the accretion rate of the NS for $a \approx 1
R_\odot$ to be $\approx 30 M_\odot \yr^{-1}$. In this regime
neutrino-cooled accretion occurs as $\dot M_{\rm acc} \ga 10^{-3}
M_\odot \yr^{-1}$ \citep{Chevalier1993}, and so high accretion
rate can take place. The Keplerian orbital time is $P_{\rm K}
\simeq 1 (a/R_\odot)^{3/2}$ hours. If evolution proceed over a
time of $10 P_{\rm K} \simeq 0.5$~day, the total accreted mass is
less than about $0.04 M_\odot$.

The total accreted mass in the jet-feedback scenario is calculated from
\begin{equation}
\frac{E_{\rm e/bind}}{\chi} \simeq  E_j \simeq \frac{\eta}{2} M_{\rm acc} v_j^2.
\label{eq:Macct1}
\end{equation}
This gives an upper limit on the accreted mass as the orbital gravitational energy released by the NS-core system reduces the required energy from the jets.
Substituting typical values used here we derive
\begin{equation}
M_{\rm acc} \la 0.04 v_{j5}^{-2}
\left(\frac{E_{\rm bind}}{4\times 10^{49} \erg } \right)
\left(\frac{\eta}{0.1}\right)^{-1} \left(\frac{\chi}{0.1}\right)^{-1} M_\odot.
\label{eq:Macct2}
\end{equation}
This is the same as the value estimated above, showing that the
final stage of the CE inside the envelope lasts for about $t_{\rm
f-CE} \simeq 1~$day. Hence, most of the accretion takes place
during the few hours to few days of the final stage of the CE,
when evolution is rapid. In the outer regions of the envelope,
that have very small binding energy and where the spiraling-in
time is of order of years, the accretion rate is very small in the
feedback scenario.

{ {{ We emphasize again that although most of the accretion energy
is taken by neutrinos, in the proposed scenario a non-negligible
fraction of the energy is carried by jets launched from the
accretion disk around the NS. When accretion rate is high, we have
$L_\nu \gg L_{\rm jets} \gg L_{\rm Edd}$, where $L_\nu$ is the
neutrino luminosity, $L_{\rm jets}$ is the jets' power, and
$L_{\rm Edd}$ is the critical Eddington luminosity from the NS.
}}}

We conclude that within the frame of the jet-feedback mechanism a
NS ends the spiraling-in inside the envelope of a giant after
accreting only a small amount of mass, that most likely leaves it
as a NS rather than forming a BH.

\subsection{Inside the core}
\label{sec:core}
We can repeat the calculations of section \ref{sec:envelope} to
the phase when the NS is inside the core. The binding energy of
the core is larger than $10^{50} \erg$. The ejected mass by the
jets in the frame of the jet-feedback mechanism is
\begin{equation}
M_{2\rm j} = \eta M_{\rm acc} \la   0.01 v_{j5}^{-2}
\left(\frac{E_{\rm bind}}{10^{50} \erg} \right)
\left(\frac{\chi}{0.1}\right)^{-1} M_\odot. \label{eq:Mjcore}
\end{equation}
The Keplerian orbital period for the core and the NS is
approximately $10$ minutes. During that time the NS can explode
the core with the jets. The explosion will last a few minutes, but
since the entire system is optically thick due to the ejected
envelope, the explosion will last days, as in a typical SN. {{{ By
explosion we refer here to the case where the jets expel the
envelope or the core within a time period much shorter than the
dynamical time of the giant, and the ejected mass has energy above
its binding energy. This occurs when the NS is deep in the
envelope or inside the core. }} }

During the NS merger with the core the Bondi-Hoyle accertion rate
can get as high as $1 M_\odot \s^{-1}$ \citep{Fryer1998}. In this
regime neutrino-cooled accretion can take place as $\dot M_{\rm
acc} \ga 10^{-3} M_\odot \yr^{-1}$ \citep{Chevalier1993}, and the
accretion rate can get near the Bondi-Hoyle rate. Simulations by
\citet{Ricker2012} show that the actual rate can be much lower
than the Bondi-Hoyle rate. In addition, \citet{Chevalier1996}
argued that angular momentum considerations can prevent neutrino
cooling from occurring.

Here we take a different approach, we assume that jets are lunched
during the NS-core merger process. These jets move relative
to the core material as the NS spirals-in and deposit part of
their energy inside the core. This limits the accretion rate
by a feedback mechanism; higher accretion rate results in higher energy deposition in the core and so suppresses the accretion process.

The NS spirals inside the core within about $5-10$ orbits, summing
up to a total interaction time of $t_j \approx 1~$hour. The
interaction of the jets with the core takes place within a radius
of $R_i \approx 0.1 R_\odot$. This accretion rate is well below
the Bondi-Hoyle rate and is comparable with the results of
\citet{Ricker2012}. The result of the process will be probably
observed as a type IIn SN \citep{Chevalier2012}.

Finally, when the NS is well inside the core, such that the core
mass that is left is about equal to the NS mass, the rest of the
core forms an accretion disk around the NS. This accretion disk of
mass $M_{\rm disk} \approx 1 M_\odot$ launches jets that interact
with gas at a larger distances of $r \approx 1 R_\odot$. This
distance comes from the distance the mass expelled from the core
at the escape speed of $v_{\rm esc} \approx 1000 \km \s^{-1}$
reaches within a fraction of an hour. Gamma Ray burst (GRBs) are
observed in Type Ic SN \citep{Woosley2006} and hence we don't
claim this system will be a GRB, since likely the jets do not
penetrate the entire envelope.

\section{IMPLICATIONS}
\label{sec:implciaitons}
{{{{{  If our claim that jets from NS can indeed remove the
envelope and part of the core of RSG stars holds, we can think of
two implications.
 Both of which require further study. }}}}}

\subsection{The red supergiant star HV2112}
\label{subsec:HV2112}
{{{{{ \cite{Levesqueetal2014} attributed the peculiar abundances
of some isotopes of the red supergiant (RSG) star HV2112 in the
Small Magellanic Cloud to the star being a Thorne-\.{Z}ytkow
object (TZO). However, some features, such as the high calcium
abundance, are not accounted for by processes occurring in TZO.
\cite{Toutetal2014} examined whether HV2112 can be a super
asymptotic giant branch (SAGB) star, a star with an oxygen/neon
core supported by electron degeneracy and undergoing thermal
pulses with a third dredge up. { {{ The initial mass range of SAGB
progenitors is thought to be $ 7 \la M_{\rm SAGB} \la 11 M_\odot$
\citep{EldridgeTout2004, Siess2006, Dohertyetal2014}, depending on
the manner convective overshooting is treated
\citep{EldridgeTout2004} and on metallicity
\citep{Dohertyetal2014}. RSG stars originate from more massive
stars than the progenitors of SAGB stars. At the evolutionary
stage of HV2112, SAGB and RSG stars occupy more or less the same
region in the HR diagram. In the TZO scenario the star HV2112 is a
RSG star, while in the scenario proposed here it is a SAGB star,
hence of a lower mass. }}}

\cite{Toutetal2014} argued that a SAGB star can synthesize most of
the elements that were used to claim HV2112 to be a TZO, e.g.,
molybdenum, rubidium, and lithium. But they found no way for a SAGB
star to synthesize calcium. They still preferred a TZO
interpretation for HV2112, and attributed the enriched calcium
envelope to the formation process of a TZO. The calcium, they
argued, can be synthesized when the degenerate electron core of a
giant star is tidally disrupted by a neutron star and forms a disk
around the NS, as in the calculations of \cite{Metzger2012} for a
WD merger with a NS. The high temperatures in such accretion
disk leads to calcium production. Interestingly, they found that
the kinetic energy of the outflow from the accretion disk that is
required to spread calcium in the giant, has enough energy to
unbind the envelope. They postulated that the outflow is
collimated, hence most of it escapes from the star. We find in
section \ref{sec:commonenvelope} that the jets from the NS are
formed while it is still in a Keplerian orbit, hence the jets are
not well collimated, and the envelope and a large part of the core
are ejected.

The peculiar abundances of some isotopes (e.g. lithium) might be
related to the presence of a binary companion. The post AGB star
HD172481, with a metallicity of [Fe/H]$=-0.55$, is in a binary
system and has a very high lithium abundance
\citep{ReyniersVanWinckel2001}. \cite{Toutetal2014} also noted
that a SAGB star can synthesize all elements, besides calcium.
They argued against the calcium enrichment scenario by a CCSN
because the fraction of intercepted ejecta by HV2112, while still
on the main sequence, is small.

Based on our calculations in section \ref{sec:commonenvelope} we
argue that a TZO cannot be formed in the process where a NS
spirals inside a RSG star. The jets launched by the accreting NS
will expel the entire envelope. We instead {{{ speculate }}} that
HV2112 had a companion just slightly more massive than HV2112 when
both were on the main sequence. In such massive binary systems the
lighter star expands to become a giant before the more massive
star explodes \citep{SabachSoker2015}. By the influence of the
companion the CCSN could have been a type Ib SN. At explosion, the
radius of HV2112, $R_2$, could have been a sizable fraction of the
orbital separation, $a$. It intercepted a fraction of $f \simeq
0.04 (R_2/0.4 a)^2$ of the SN ejecta. We use this ratio to replace
equation (4) of \cite{Toutetal2014} where the factor is scaled to
$ f \simeq 4 \times 10^{-4}$. If, for example, the companion's
initial mass was about $12 M_\odot$, it could have ejected
$\approx 0.01 M_\odot$ of calcium \citep{WoosleyWeaver1995}. As
the mass of calcium in HV2112 is estimated by \cite{Toutetal2014}
to be $\approx 10^{-4} M_\odot$, it requires about a quarter of
the SN ejecta to hit HV2112 during the explosion and stays bound
to it, for such an orbital separation (for more detail see
\citealt{SabachSoker2015}). After the explosion the NS was kicked
out of the binary system. }}}}}

{{{ The main question for the proposed scenario is whether indeed
such a large amount of calcium can be accreted to the companion.
While some other works suggested this possibility for other
systems, e.g., for the hyper--runaway star HD271791
\citep{Schaffenrothetal2014}, recent numerical simulations by
\cite{Hiraietal2014} suggested that this is not possible. The SN
ejecta that encounter the companion induce a shock that runs
through the companion. The shock heats the companion and the
excess energy leads to mass removal. In their numerical
simulations they found that up to $25 \%$ of the companion mass
can be removed; this is the case when the companion is as close as
possible, as the scaling we used above of $R_2 \simeq 0.4 a$. If
this is indeed the case, then our proposed scenario cannot work.
}}}

{{{ However, there are some processes that might change the
outcome, in particular non-spherical SN ejecta of calcium (and
other newly synthesized elements). A non spherical ejection of
synthesized elements is expected in the jittering-jets model for
core collapse SN explosion \citep{Papishsoker2014,
Papishsoker2014b}. One possibility is that the amount of calcium
produced in the SN explosion is larger, or at least the calcium
distribution is non-spherical, with calcium-rich gas ejected
toward the companion. Another process that can overcome the
difficulties posed by the results of \cite{Hiraietal2014} and
allows large quantities of calcium and other heavy elements to be
accreted on to the companion is if the newly synthesized elements
from the core of the SN expand in dense clumps. Such clumps can
penetrate deeper to the star, and stay bound. We consider the
question of whether SN ejecta can enriched a companion as not
settled yet, but it is the weak point of the speculative scenario
proposed here. }}}

\subsection{The r-process}
\label{subsec:rprocess}

Our analysis in section \ref{sec:commonenvelope} points to three
types of interactions of the NS jets inside a common envelope,
CE-SN jets, with a star that could in principle take place. ($i$)
Small amounts of jets' mass interact with the low density envelope
during the final stages of the inspiral inside the envelope.
($ii$) Next a jets' mass of about $0.01 M_\odot$ interacts with
the dense core when the NS enters the core. Interaction occurs
within a distance of about $0.1 R_\odot$. The amount of mass
carried by the jets and their typical distance of interaction with
stellar mater in the above two types of jets, is determined by the
assumptions of the jet-feedback mechanism. One has to bear in mind
the large uncertainties. ($iii$) Finally, the central leftover of
the core is destructed by the NS gravity and an accretion disk of
about $1M_\odot$ is formed around the NS. This disk is expected to
launch jets with a total mass of about $0.1 M_\odot$, that
catch-up with previously ejected core material at $\approx 1
R_\odot$. The last two stages occur within a few minutes, and the
last launching episode, the most massive one, lasts for tens to
hundreds of seconds.

The R-process can generally occur both in the jets
\citep{Cameron2001,Winteleretal2012} and in the post shock jets'
material  \citep{PapishSoker2012}. In the type $(i)$ jet
interaction listed above the amount of jet material is small and
its contribution to the r-process can be neglected. For the latter
phases the post shock temperature can be estimated from the post
shock radiation dominated pressure, $P = \frac{6}{7}\frac{\dot M_j
v_j}{4\uppi \delta r^2}$
\begin{eqnarray}
&T \simeq 2\times 10^8 \left( \frac{\dot M_j}{10^{-3} M_\odot/ \s}\right)^{1/4}
\left( \frac{v_j}{10^5 \km \s^{-1}}\right)^{1/4} \left( \frac{r}{R_\odot}\right)^{-1/2}  \nonumber
\\
&\times \left( \frac{\delta}{0.1}\right)^{-1/4} \K.
\label{eq:temp}
\end{eqnarray}
Here $\delta$ is the solid opening angle of the jet and $r$ is the
radius where the jet is shocked. This is much too low for
r-process to occur in the post shock jets' material at a distance
of $r \approx 0.1-1 R_\odot$. In these cases r-process will occur
inside the jets.

{{{ The flow structure studied here is different in key
ingredients from two other types of jet launching models in CCSNe.
In the jittering jets model \citep{Papishsoker2014}, listed in the
second column of Table \ref{table:jets},  jets are launched by a
newly formed NS. However, each launching episode lasts for a short
time of less than $0.1 \s$, and the disk formed at each launching
episode is short lived. It is not clear that the neutron fraction
in the ejecta will be as high as in the case studied here, where
the magnetic fields lift material from very close to the NS
\citep{Winteleretal2012}. Another difference is that the high
neutrino luminosity from the newly formed NS in the jittering jets
model is likely to bring the ejecta closer to equilibrium, e.g.,
lower the neutron fraction. A third difference is that the
jittering jets are expected to explode the star, and so they are
shocked relatively close to the center at $r_{\rm sh} \approx
0.001-0.01 R_\odot$ \citep{PapishSoker2012}. Even if strong-r
process elements have been synthesize in the jets, they will be
disintegrated in the shock. In the flow studied in the present
paper the shock is expected to take place much further out, and
the core of the primary star has been already destroyed and formed
the massive accretion disk now circling the NS.  }}}

{{{ Our flow structure is markedly different from the jets
launched in GRB that are formed around black holes (BH). In the
flow studied by \cite{Surnametal2006}, for example, the jets are
launched at $r=100$ and $250 \km$ from the center, compared with
$r \approx 15 \km $ in the present study, and the entropy
considered is $s/k=10-50$, higher than in the study of
\cite{Winteleretal2012}. The same holds for the neutrino wind in
the study of \cite{Pruet2005}, that within a very short time
reaches an entropy per baryon of $s/k=50-80$. In both
\cite{Surnametal2006} and \cite{Pruet2005} the outflow starts as
neutron-rich, $Y_e \approx 0.2$, but the high entropy implies that
positron convert neutrons to protons. }}}
\begin{table}[h!]
\begin{tabularx}{ 1\textwidth}{ X X X X X}
    \hline
    ~                   & CCSN jittering jets                              & CCSN MHD jets &  NS common-envelope jets & GRB jets
    \\ \hline
    source & \cite{PapishSoker2012} & \cite{Winteleretal2012} & This Paper & \citet{Popham1999,Pruet2005} \\ \hline \hline
    Activity duration ($s$)          & $\approx 1$           & $<1$                  & $\approx 100$  & $\approx 1-1000$ \\
    Ejected mass $(M_\odot)$            & $\approx 0.01$           &               $\approx 0.01$  & $\approx 0.01-0.1$  & $\approx 10^{-5} - 10^{-6}$       \\
    Shock distance $(R_\odot)$            & $\approx 0.001-0.01$          &---               & $\approx 0.1-1$  & ---\\
    $L_{\nu} (\erg \s^{-1})$ & $\approx 5 \times 10^{52}$         &     $\approx 5 \times 10^{52}$            & $\la 10^{50} $ & $\approx 0.01 - 1000 \times 10^{51}$ \\
    Compact Object & NS & NS & NS & BH \\
    $Y_e$ & $\approx 0.15$ & $\approx 0.15$ & assumed to be as in \cite{Winteleretal2012} & $\approx 0.5$ for low accretion rate \\

    Frequency and r-process nucleosynthesis             & Weak r-process in most CCSNe. &  Rapid core rotation in 1\% of CCSNe. Strong r-process takes place.
 & ~1\% relative to all CCSNe. Strong r-process takes place.  &  \\
    \hline
     \end{tabularx}
    \caption{Summery of the differences between CCNS jets, CE-NS jets and GRBs jets that provide possible sites for r-process nucleosynthesis.
    The properties of the jets in CCSNe are based on the jitering-jets model (JJM; \citealt{papish2011}), or the magnetic-jets studied by
    \citet{Winteleretal2012}.
    The number of CE-NS mergers relative to all CCSNe is taken from \citet{Chevalier2012}. Data for GRBs is taken from \citet{Popham1999,Pruet2005}. {{{ Only in the flow structure of CCSN MHD jets and the NS-CE jets the accretion disk is both steady and bounded from inside, by the NS surface. These lead to low entropy neutron-rich outflow that facilitate the nucleosynthesis of strong r-process (third peak) elements.   }}}}
    \label{table:jets}
\end{table}

{{{ In CCNSe the high neutrino luminosity of $L_\nu \approx
5\times 10^{52} \erg \s^{-1}$ can suppress the r-process by
raising the electron fraction (reducing the neutron fraction)
through weak interactions (e.g., \citealt{Pruet2006,
Fischer2010}). This has a large effect on the formation of
r-process elements in neutrino driven winds and could affect to
some degree the r-process in CCNSe jets \citep{Winteleretal2012}.
In particular, high neutrino luminosity can suppresses the third
peak of the r-process. The synthesis of the third peak, the strong
r-process, is a rare process relative to the synthesis process of
the first two peaks, as evident from the large abundance
variations of Eu/Fe found in old stars \citep{Cowan2011}.
\cite{Winteleretal2012} found that in magnetic-jet simulation
strong r-process with mass of about $6 \times 10^{-3} M_\odot$ is
synthesized. To account for the third peak abundance they required
that one in about $100$ CCSNe has the strong r-process.

{\citet{Fryer2006} studied the ejection of matter from a supernova
fallback as a site for the r-process. They found that the
conditions in the ejecta are compatible with the production of the
'strong' r-process for accretion rate similar to those in our
proposed  scenario,$\dot M_{\rm acc} \approx 3\times 10^{-4}
M_\odot \s^{-1}$ and have an initial electron fraction of $Y_e =
0.5$. In their calculations the ejecta is driven by energy
released during the accretion on to the surface of the NS, but not
in a jet like driven outflow. } }}}

The CE-NS jets can have the required properties to account for the
strong r-process. First, based on \cite{Podsiadlowski1995},
\cite{Chevalier2012} estimated that about $1 \%$ of observed CCSNe
are from a NS-core merger. The ejected mass in the jets in our
proposed CE-NS scenario is $0.01-0.1 M_\odot$, which can lead to a
synthesize of approximately $0.001-0.01 M_\odot$ of heavy
elements. This ratio is based on the simulations of jets in CCSNe
performed by \cite{Nishimura2006} who found that the total amount
of r-process elements is approximately $10\%$ of the total jets
ejected mass. This mass is similar to what \cite{Winteleretal2012}
have found.

\section{SUMMARY}
\label{sec:summary}

We studied the common-envelope (CE) evolution of a NS inside a red
supergiant (RSG; section \ref{sec:commonenvelope}), and found that
the NS is very likely to launch energetic jets that might expel
the entire envelope and most of the core of the giant star. Based
on this, we proposed that Thorne--\.Zytkow objects are unlikely to
be formed by this channel if accretion on to the neutron star can
exceed the Eddington rate with much of the accretion energy
directed into jets that subsequently dissipate within the giant
envelope. We then discussed the implications of our results to the
recent claim that the evolved star HV2112 is a TZO
\citep{Levesqueetal2014, Toutetal2014}, and to r-process
nucleosynthesis inside the jets.

{{{ In the TZO model for HV2112 \citep{Levesqueetal2014,
Toutetal2014} the star is a RSG star originated from a main
sequence star with an initial mass of $M_{\rm MS}(\rm TZO)\approx
15 M_\odot$. }}}  In section \ref{subsec:HV2112} we proposed an
alternative scenario for the peculiar abundances of HV2112. {{{ In
the proposed rare scenario HV2112 is presently a super asymptotic
giant branch (SAGB) star that originated from a main sequence star
with an initial mass of $M_{\rm MS}(\rm SAGB)\approx 8.5-11
M_\odot$. }}} Beside calcium, SAGB can synthesize all elements
with high abundances \citep{Toutetal2014}. We suggest that HV2112
had a companion slightly more massive than the initial mass of
HV2112. The companion evolved first, but it exploded when HV2112
was already a giant \citep{SabachSoker2015}. This implies that
HV2112 intercepted a large fraction of the supernova ejecta,
including a sufficient mass of calcium. {{{ We also point out some
weak points in our proposed scenario. In particular the question
of whether a giant companion can accrete a large enough fraction
from the SN ejecta that hits it. It seems that our proposed
scenario might work only if the newly synthesized elements are
ejected from the core of the SN in non-spherical structures, i.e.,
clumps or jets toward the companion. }}}

In section \ref{subsec:rprocess} we discussed the possibility that
jets launched by the NS as it merges with the giant's core could
be a rare site for the strong r-process. Although most ingredients
of the proposed scenario were studied in the past, they were never
put together to yield the scenario we have proposed in this study.
For example, the NS-core merger was studied in the past as a
possible candidate for supernovae events where strong interaction
with a dens environment takes place \citep{Chevalier2012}. As the
NS accretes mass from the giant at a very high rate due to
neutrino cooling, an accretion disk is formed around the NS and
two opposite jets are launched during the spiral-in process
\citep{Armitage2000}. We estimated the total energy deposited by
the jets inside the envelope (eq. \ref{eq:ej}) using the
jet-feedback mechanism \citep{Sokeretal2013}, where energy deposit
by the jets regulates the accretion rate. A small amount of
accreted mass is sufficient to launch jets that expel the envelope
(section \ref{sec:envelope}).

In many cases the NS merges with the core. We termed the jets
launched by the NS during its interaction with the core `CE-NS
jets'. In these cases we have found that much stronger jets are
launched (section \ref{sec:core}), unbind, and expel the outer
parts of the core. The total mass launched by the jets in the
outer part of the core is about $0.01M_\odot$. The jets in this
phase interact with the surrounding core matter at a distance of
$r \approx 0.1R_\odot$. This distance is too large for the
r-process to occur inside the post shock jets' material due to low
post-shock temperature (eqaution \ref{eq:temp}). However, the
r-process can take place inside the jets close to the center
before they are shocked, much as in the MHD jets studied by
\citet{Winteleretal2012}.

In the third and final phase when the NS is well inside the core
an accretion disk with a mass of about $1M_\odot$ is form around
the NS from the destructed core. This accretion disk launches two
opposite jets with a total mass of approximately $0.1M_\odot$. We
find that in total $0.01-0.1 M_\odot$ of jets' material is ejected
by the NS as it interacts with the core (see Table
\ref{table:jets}). Nucleosynthesis of r-process elements can take
place inside these jets \citep{Cameron2001} with a total r-process
material that can be as high as about $0.01 M_\odot$.

We compared the properties of the CE-SN jets with jets launched in
CCSNe and found that the flow studied here is similar to that of
MHD jets launched by newly formed NS when the pre-collapse core is
rapidly rotating \citep{Winteleretal2012}.
\citet{Winteleretal2012} showed that under these conditions of
low-entropy neutron-rich outflow the third r-process peak elements
can be synthesized (the strong r-process). Lower neutrino luminosity
of the NS in the NS-core merger (Table \ref{table:jets})  favors
the production of strong r-process in the CE-NS jets compared to
CCSNe jets. The rareness of this process of approximately $1\%$ of
the CCSNe rate \citep{Chevalier2012} can  explain the large
scattering of r-process elements in the early chemical evolution
of the galaxy (\citealt{Argast2004, Winteleretal2012}).

{{{ We thank Christopher Tout, the referee, for very valuable and
detail comments. }}}
 Oded Papish thanks Friedrich-Karl Thielemann for his hospitality and
discussions of the r-process.



\begin{thebibliography}{}

\bibitem[Argast et al.(2004)]{Argast2004} Argast, D., Samland, M., Thielemann, F.-K., \& Qian, Y.-Z.\ 2004, \aap, 416, 997

\bibitem[Armitage \& Livio(2000)]{Armitage2000} Armitage, P.~J., \& Livio, M.\ 2000, \apj, 532, 540

\bibitem[Bisnovatyi-Kogan et al.(1976)]{Bisnovatyi1976} Bisnovatyi-Kogan, G.~S., Popov, I.~P., \& Samokhin, A.~A.\ 1976, \apss, 41, 287

\bibitem[Bondi \& Hoyle(1944)]{BondiHoyle1944} {{{ Bondi, H., \& Hoyle, F.\
1944, \mnras, 104, 273 }}}

\bibitem[Cameron(2001)]{Cameron2001} Cameron, A.~G.~W.\ 2001, \apj, 562, 456

\bibitem[Chevalier(1993)]{Chevalier1993} Chevalier, R.~A.\ 1993, \apjl, 411, L33

\bibitem[Chevalier(1996)]{Chevalier1996} Chevalier, R.~A.\ 1996, \apj, 459, 322

\bibitem[Chevalier(2012)]{Chevalier2012} Chevalier, R.~A.\ 2012, \apjl, 752, L2

\bibitem[Couch et al.(2009)]{Couch2009} Couch, S.~M., Wheeler, J.~C., \& Milosavljevi{\'c}, M.\ 2009, \apj, 696, 953

\bibitem[Cowan et al.(2011)]{Cowan2011} Cowan, J.~J., Roederer, I.~U., Sneden, C., \& Lawler, J.~E.\ 2011, RR Lyrae Stars, Metal-Poor Stars, and the Galaxy, 223

\bibitem[De Marco et al.(2012)]{DeMarcoetal2012}{ {{  De Marco, O., Passy,
J.-C., Herwig, F., Fryer, C. L., Mac Low, M.-M., \& Oishi, J. S.\
2012, IAU Symposium, 282, 517 }}}

\bibitem[Doherty et al.(2014)]{Dohertyetal2014} {{{ Doherty, C.~L.,
Gil-Pons, P., Siess, L., Lattanzio, J.~C., \& Lau, H.~H.~B.\ 2014,
arXiv:1410.5431  }}}

\bibitem[Eldridge \& Tout(2004)]{EldridgeTout2004} {{{ Eldridge, J.~J., \& Tout,
C.~A.\ 2004, \memsai, 75, 694 }}}

\bibitem[Fischer et al.(2010)]{Fischer2010} Fischer, T., Whitehouse, S.~C., Mezzacappa, A., Thielemann, F.-K., \& Liebend{\"o}rfer, M.\ 2010, \aap, 517, A80

\bibitem[Fryer et al.(1996)]{Fryer1996} Fryer, C.~L., Benz, W., \& Herant, M.\ 1996, \apj, 460, 801

\bibitem[Fryer et al.(2006)]{Fryer2006} Fryer, C.~L., Herwig, F., Hungerford, A., \& Timmes, F.~X.\ 2006, \apjl, 646, L131

\bibitem[Fryer \& Woosley(1998)]{Fryer1998} Fryer, C.~L., \& Woosley, S.~E.\ 1998, \apjl, 502, L9

\bibitem[Gilkis \& Soker(2014)]{GilkisSoker2014} {{{ Gilkis, A., \& Soker, N.\
2014, \mnras, 439, 4011 }}}

\bibitem[Hirai et al.(2014)]{Hiraietal2014} {{{ Hirai, R., Sawai, H.,
\& Yamada, S.\ 2014, \apj, 792, 66  }}}

\bibitem[Hoffman et al.(1997)]{Hoffman1997} Hoffman, R.~D., Woosley, S.~E., \& Qian, Y.-Z.\ 1997, \apj, 482, 951

\bibitem[Houck \& Chevalier(1991)]{HouckChevalier1991} Houck, J.~C., \& Chevalier, R.~A.\ 1991, \apj, 376, 234

\bibitem[Hoyle \& Lyttleton(1939)]{HoyleLyttleton1939} {{{ Hoyle, F., \& Lyttleton,
R.~A.\ 1939, Proceedings of the Cambridge Philosophical Society,
35, 405  }}}

\bibitem[Iben \& Livio(1993)]{Iben1993} Iben, I., Jr., \& Livio, M.\ 1993, \pasp, 105, 1373

\bibitem[Ivanova et  al.(2013)]{Ivanovaetal2013} Ivanova, N., Justham, S., Chen, X., et al.\ 2013, \aapr, 21, 59

\bibitem[Khokhlov et al.(1999)]{Khokhlov1999} Khokhlov, A.~M., H{\"o}flich, P.~A., Oran, E.~S., et al.\ 1999, \apjl, 524, L107

\bibitem[Kohri et al.(2005)]{kohri2005} Kohri, K., Narayan, R., \& Piran, T.\ 2005, \apj, 629, 341

\bibitem[LeBlanc \& Wilson(1970)]{LeBlanc1970} LeBlanc, J.~M., \& Wilson, J.~R.\ 1970, \apj, 161, 541

\bibitem[Levesque et al.(2014)]{Levesqueetal2014} Levesque, E.~M.,
Massey, P., {\.Z}ytkow, A.~N., \& Morrell, N.\ 2014, \mnras, 443, L94

\bibitem[MacFadyen et al.(2001)]{MacFadyen2001} MacFadyen, A.~I., Woosley, S.~E., \& Heger, A.\ 2001, \apj, 550, 410

\bibitem[Meier et al.(1976)]{Meier1976} Meier, D.~L., Epstein, R.~I., Arnett, W.~D., \& Schramm, D.~N.\ 1976, \apj, 204, 869

\bibitem[Metzger(2012)]{Metzger2012} Metzger, B.~D.\ 2012, \mnras, 419, 827

\bibitem[Nishimura et al.(2006)]{Nishimura2006} Nishimura, S., Kotake, K., Hashimoto, M.-a., Yamada, S., Nishimura, N., Fujimoto, S., \& Sato, K.\ 2006, \apj, 642, 410

\bibitem[Papish \& Soker(2011)]{papish2011} Papish, O., \& Soker, N.\ 2011, \mnras, 416, 1697

\bibitem[Papish \& Soker(2012)]{PapishSoker2012} Papish, O., \& Soker, N.\ 2012, \mnras, 421, 2763

\bibitem[Papish \& Soker(2014a)]{Papishsoker2014} Papish, O., \& Soker, N.\ 2014a, \mnras, 438, 1027

\bibitem[Papish \& Soker(2014b)]{Papishsoker2014b} {{{ Papish, O., \& Soker, N.\
2014b, \mnras, 443, 664 }}}

\bibitem[Passy et al.(2012)]{Passyetal2012} {{{ Passy, J.-C., De Marco,
O., Fryer, C.~L., et al.\ 2012, \apj, 744, 52 }}}

\bibitem[Podsiadlowski et al.(1995)]{Podsiadlowski1995} Podsiadlowski, P., Cannon, R.~C., \& Rees, M.~J.\ 1995, \mnras, 274, 485

\bibitem[Popham et al.(1999)]{Popham1999} Popham, R., Woosley, S.~E., \& Fryer, C.\ 1999, \apj, 518, 356

\bibitem[Pruet et al.(2005)]{Pruet2005} Pruet, J., Woosley, S.~E., Buras, R., Janka, H.-T., \& Hoffman, R.~D.\ 2005, \apj, 623, 325

\bibitem[Pruet et al.(2006)]{Pruet2006} Pruet, J., Hoffman,
R.~D., Woosley, S.~E., Janka, H.-T., \& Buras, R.\ 2006, \apj, 644, 1028

\bibitem[Reyniers \& Van Winckel(2001)]{ReyniersVanWinckel2001} {{{{{ Reyniers, M., \& Van Winckel, H.\ 2001, \aap, 365, 465 }}}}}

\bibitem[Schaffenroth et al.(2014)]{Schaffenrothetal2014} {{{ Schaffenroth, V.,  Przybilla, N., Butler, K., \& Heber, U. 2014,
Binary Systems, their Evolution and Environments, P1P }}}

\bibitem[Surman et al.(2006)]{Surnametal2006} Surman, R., McLaughlin, G.~C., \& Hix, W.~R.\ 2006, \apj, 643, 1057

\bibitem[Qian(2000)]{Qian2000} Qian, Y.-Z.\ 2000, \apjl, 534, L67

\bibitem[Qian(2012)]{Qian2012} Qian, Y.-Z.\ 2012, American Institute of Physics Conference Series, 1484, 201


\bibitem[Ricker \& Taam(2012)]{Ricker2012} Ricker, P.~M., \& Taam, R.~E.\ 2012, \apj, 746, 74

\bibitem[Rosswog et al.(2014)]{Rosswogetal2013} Rosswog, S., Korobkin,
O., Arcones, A., Thielemann, F.-K., \& Piran, T.\ 2014, \mnras, 439, 744

\bibitem[Sabach \& Soker(2015)]{SabachSoker2015} {{{{{  Sabach, E., \& Soker, N.\ 2014, arXiv:1410.1713   }}}}}

\bibitem[Siess(2006)]{Siess2006} {{{ Siess, L.\ 2006, \aap, 448,
717 }}}

\bibitem[Sneden et al.(2008)]{Sneden2008} Sneden, C., Cowan, J.~J., \& Gallino, R.\ 2008, \araa, 46, 241

\bibitem[Soker(2004)]{Soker2004} Soker, N.\ 2004, \na, 9, 399

\bibitem[Soker(2013)]{Soker2013} Soker, N.\ 2013, \na, 18, 18

\bibitem[Soker(2014)]{Soker2014} {{{{{ Soker, N.\ 2014, arXiv:1404.5234 }}}}}

\bibitem[Soker(2015)]{Soker2015} {{{  Soker, N.\ 2015, arXiv:1410.5363
}} }
\bibitem[Soker et al.(2013)]{Sokeretal2013} Soker, N., Akashi, M., Gilkis, A.,  Hillel, S., Papish, O., Refaelovich, M., Tsebrenko, D.\ 2013, Astronomische Nachrichten, 334, 402

\bibitem[Surman \& McLaughlin(2004)]{Surman2004} Surman, R., \& McLaughlin, G.~C.\ 2004, \apj, 603, 611

\bibitem[Taam et al.(1978)]{Taam1978} Taam, R.~E., Bodenheimer, P., \& Ostriker, J.~P.\ 1978, \apj, 222, 269

\bibitem[Thielemann et al.(2011)]{Thielemannetal2011} Thielemann, F.-K., Arcones, A., K{\"a}ppeli, R., et al.\ 2011, Progress in Particle and Nuclear Physics, 66, 346

\bibitem[Thorne \& Zytkow(1975)]{ThorneZytkow1975} Thorne, K.~S., \& Zytkow, A.~N.\ 1975, \apjl, 199, L19

\bibitem[Tout et al.(2014)]{Toutetal2014} {{{{{  Tout, C.~A., Zytkow, A.~N., Church, R.~P., \& Lau, H.~H.~B.\ 2014, arXiv:1406.6064 }}}}}

\bibitem[Winteler et al.(2012)]{Winteleretal2012} Winteler, C., K{\"a}ppeli, R., Perego, A., Arcones, A., Vasset, N., Nishimura, N., Liebendï¿½rfer, M., Thielemann, F.-K.\ 2012, \apjl, 750, L22

\bibitem[Woosley \& Janka(2005)]{Woosley2005} Woosley, S., \& Janka, T.\ 2005, Nature Physics, 1, 147

\bibitem[Woosley \& Bloom(2006)]{Woosley2006} Woosley, S.~E., \& Bloom, J.~S.\ 2006, \araa, 44, 507

\bibitem[Woosley \& Weaver(1995)]{WoosleyWeaver1995} Woosley, S.~E., \& Weaver, T.~A.\ 1995, \apjs, 101, 181

\end{thebibliography}
\end{document}